\documentclass[10pt,preprint]{aastex}

\begin{document}

\shortauthors{Luhman et al.}
\shorttitle{New Taurus Members}

\title{New Low-Mass Members of the Taurus Star-Forming Region
\altaffilmark{1}}

\author{K. L. Luhman\altaffilmark{2}, C\'esar Brice\~no\altaffilmark{3}, 
John R. Stauffer\altaffilmark{4}, Lee Hartmann\altaffilmark{5},
D. Barrado y Navascu\'es\altaffilmark{6}, and Nelson Caldwell\altaffilmark{7}}

\altaffiltext{1}{Based on observations obtained at the Kitt Peak National 
Observatory, Keck Observatory, Fred Lawrence Whipple Observatory, and the MMT 
Observatory.
This publication makes use of data products from the Two Micron All  
Sky Survey, which is a joint project of the University of Massachusetts 
and the Infrared Processing and Analysis Center/California Institute 
of Technology, funded by the National Aeronautics and Space
Administration and the National Science Foundation.}

\altaffiltext{2}{Harvard-Smithsonian Center for Astrophysics, 60 Garden St., 
Cambridge, MA 02138, USA; kluhman@cfa.harvard.edu.}

\altaffiltext{3}{Centro de Investigaciones de Astronom{\'\i}a,
Apartado Postal 264, M\'erida 5101-A, Venezuela; briceno@cida.ve.}

\altaffiltext{4}{SIRTF Science Center, Caltech MS 314-6, Pasadena, CA 91125, 
USA; stauffer@ipac.caltech.edu.}

\altaffiltext{5}{Harvard-Smithsonian Center for Astrophysics, 60 Garden St., 
Cambridge, MA 02138, USA; lhartmann@cfa.harvard.edu.}

\altaffiltext{6}{Laboratorio de Astrof{\'\i}sica Espacial y F{\'\i}sica 
Fundamental, INTA, PO Box 50727, 28080 Madrid, Spain; barrado@laeff.esa.es.}

\altaffiltext{7}{Harvard-Smithsonian Center for Astrophysics, 60 Garden St., 
Cambridge, MA 02138, USA; ncaldwell@cfa.harvard.edu.}

\begin{abstract}

Brice\~no et al.\ recently used optical imaging, data from the Two-Micron 
All-Sky Survey (2MASS), and follow-up spectroscopy to search for young low-mass 
stars and brown dwarfs in 8~deg$^2$ of the Taurus star-forming region. 
By the end of that study, there remained candidate members of Taurus that 
lacked the spectroscopic observations needed to measure spectral types and 
determine membership. In this work, we have obtained spectroscopy of the 22 
candidates that have $A_V\leq8$, from which we find six new Taurus 
members with spectral types of M2.75 through M9. 
The new M9 source has the second latest spectral type of the 
known members of Taurus ($\sim0.02$~$M_\odot$). Its spectrum contains
extremely strong emission in H$\alpha$ ($W_{\lambda}\sim950$~\AA) as well 
as emission in He~I 6678~\AA\ and the Ca~II IR triplet. This is the least 
massive object known to exhibit emission in He~I and Ca~II, which together
with the strong H$\alpha$ are suggestive of intense accretion.

\end{abstract}


\section{Introduction}

Magnitude-limited searches for members of nearby star-forming regions are an 
important basis for a variety of studies of young stars and brown dwarfs.
For instance, such surveys provide well-defined membership samples from which 
unbiased initial mass functions (IMFs) can be derived. 
In addition, newly discovered young brown dwarfs are valuable targets for
detailed studies of the formation of objects at the bottom of the mass function.

The Taurus dark cloud complex has proven to be amenable to a census of its
low-mass stars and brown dwarfs.
The members of this region are nearby (140~pc) and
young (1~Myr), and thus can be detected down to very low masses. In addition,
because a majority of the members of Taurus (excluding protostars) exhibit
relatively low extinction ($A_V\lesssim5$), they are accessible to measurements 
at both optical and infrared (IR) wavelengths. Over time, surveys for new 
members in this low-density, dispersed population ($n\sim 1$-10~pc$^{-3}$) have
steadily expanded in areal coverage \citep{ss94,bri98,lr98,luh00,mar01}. 
Most recently, \citet{bri02} used wide-field optical imaging,
2MASS data, and followup spectroscopy to search for low-mass stars and brown 
dwarfs in 8~deg$^2$ that encompassed half of the known pre-main sequence 
population in Taurus.
They discovered nine new members with spectral types of M5.75-M9.5, 
corresponding to masses of 0.1-0.015~$M_\odot$ by the theoretical evolutionary 
models of \citet{bar98}.

At the conclusion of the study by \citet{bri02}, there remained
candidate members of Taurus that had not been observed spectroscopically.
In this paper, we identify the most promising of those candidates
(\S~\ref{sec:cand}),
describe spectroscopic observations of them (\S~\ref{sec:obs}), measure spectral
types and determine whether they are Taurus members (\S~\ref{sec:class}),
and discuss the physical properties of the six new members and update 
the IMF from \citet{bri02} (\S~\ref{sec:prop}).

\section{Selection of Candidate Members of Taurus}
\label{sec:cand}

\citet{bri02} used optical and near-IR photometry to select potential
young stars and brown dwarfs in 8~deg$^2$ of Taurus.
In this section, we examine the photometry from that survey to identify the
remaining candidates that lack the spectroscopy needed to
confirm membership in Taurus and to measure spectral types.
To do this, we could simply use the observed 
color-magnitude and color-color diagrams (e.g., $I$ vs.\ $I-Z$, 
$J-H$ vs.\ $I-K_s$) in the manner described by \citet{bri02}.
Instead, we revise those methods in a way that more clearly separates 
Taurus members from field stars.

\begin{figure}
\plotone{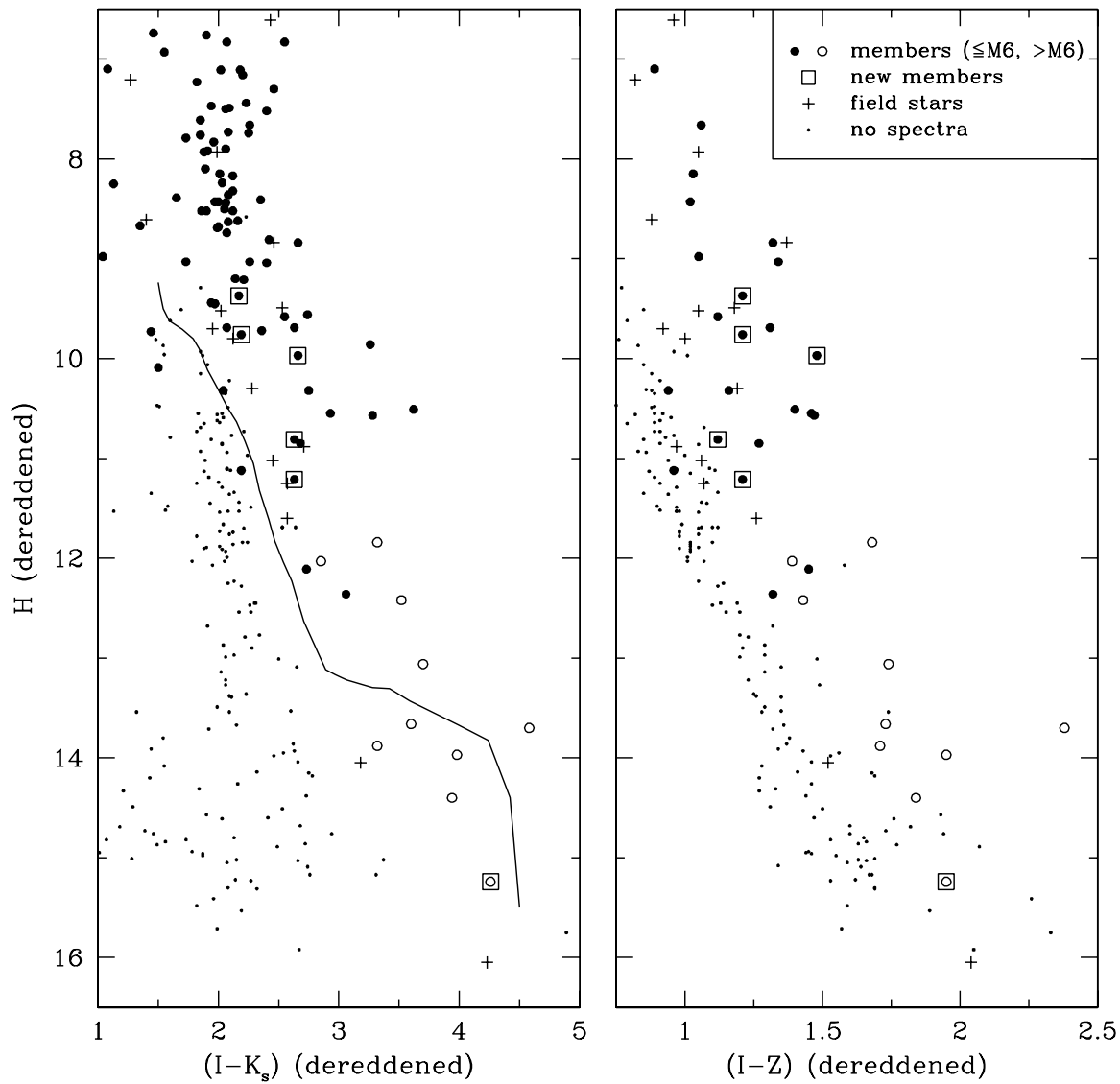}
\caption{
Extinction-corrected color-magnitude diagrams for stars with $A_V\leq8$ in the 
8~deg$^2$ surveyed by \citet{bri02} in the Taurus star-forming region. 
We plot the stars that have been spectroscopically confirmed as
Taurus members at $\leq$M6 and $>$M6 ({\it large points and open circles})
and indicate the ones found in this work ({\it boxes}).
Members later than M6 are likely to be brown dwarfs by the H-R diagram and
evolutionary models in Figure~\ref{fig:hr}.
The field stars identified in this study are also shown ({\it plusses}). 
The remaining stars lack spectra ({\it small points}). The solid line is the 
10~Myr isochrone (1-0.015~$M_{\odot}$) from the evolutionary models of 
\citet{bar98}. The stars below the isochrone at $H=9.5$-11.5 are 
known or suspected to be seen in scattered light \citep{bri02}.
}
\label{fig:color}
\end{figure}

We wish to identify the stars in the fields surveyed by \citet{bri02}
that have both the colors and the absolute photometry expected of 
members of the Taurus population. Optical color-magnitude diagrams (CMDs)
are widely used for selecting candidate members of clusters on this basis.
However, in star-forming clouds, the presence of highly variable extinction 
($A_V=0$-20 in Taurus) increases the contamination of background stars
in the region of the CMD inhabited by members and prevents
reliable mass estimates for candidate members (i.e., reddened high-mass members
and unreddened low-mass members can have the same optical color and magnitude). 
The photometry in such diagrams cannot be corrected for extinction when
only two bands are available. In the case of the Taurus fields, in addition 
to $I$ and $Z$ photometry, we have near-IR data from 2MASS.
As a result, we have been able to estimate the extinction toward each star
by dereddening its position in $J-H$ versus $I-K_s$ until it intersects the 
sequence of colors for dwarfs later than K6 \citep{leg92}, which is the range
of spectral types occupied by most known members of Taurus.
If a star has a spectral type earlier than K6, the extinction will be 
underestimated in this method, but that is acceptable. The dereddening process 
tends to remove background stars from the area of the CMD where 
members reside, and if a star's extinction is underestimated, then at worst
it remains in that area and is selected as a candidate. In other words, an 
underestimate of extinction will not result in the rejection of a 
bonafide member. After correcting the photometry for extinction in this way, 
we plot the results on CMDs of $I-K_s$ versus $H$ and $I-Z$ versus $H$
in Figure~\ref{fig:color}.
We choose $H$ as the magnitude because it correlates closely with bolometric
luminosity, requires only a small correction for extinction, and is not 
susceptible to significant contamination from excess emission from 
circumstellar material.
Meanwhile, the $I-K_s$ and $I-Z$ colors are used because they increase rapidly
with decreasing mass, producing a member sequence that is distinct from the
location of most field stars in the CMDs.
In Figure~\ref{fig:color}, we omit the stars appearing below the 
reddening vector in the diagram of $I$ versus $I-Z$ in \citet{bri02} because
they are well below the member sequence and thus are likely to be field stars.
From the $\sim250$ stars that remain, we plot only those that exhibit 
extinctions of $A_V\leq8$. We are primarily interested in 
identifying candidate members at lower extinctions that might fall
within the limit of $A_V\leq4$ that defined the IMF from \citet{bri02}
so that we can approach 100\% completeness for that sample. The extinctions
derived for the construction of Figure~\ref{fig:color} are only rough estimates 
since they are based on photometry alone without any information on the 
spectral types. Therefore, to be certain that members at $A_V\leq4$ were not 
missed, we considered a threshold of $A_V\leq8$ in selecting candidates.
In Figure~\ref{fig:color}, we include all objects that have been 
spectroscopically confirmed as members in this work and in
previous studies for the Taurus fields in question. We also indicate the field
stars identified through our new spectroscopy, while omitting the field stars
found in previous work.

We now describe the regions in Figure~\ref{fig:color} that are inhabited by 
Taurus members and examine the photometry to identify any potential members 
that lack previous spectra. 
When the known members of Taurus are placed on a Hertzsprung-Russell (H-R)
diagram, they exhibit median and maximum ages of $\sim1$ and 10~Myr, 
respectively, on the model isochrones of \citet{bar98}, except at 
the latest spectral types, where some of the members appear between the 
isochrones for 10 and 30~Myr (\citet{bri02}, \S~\ref{sec:ext}). 
We plot the 10~Myr isochrone from 0.015 to 1~$M_{\odot}$ in the diagram of
$I-K_s$ versus $H$ in Figure~\ref{fig:color} by combining the predicted
effective temperatures and bolometric luminosities \citep{bar98}, a temperature 
scale this is compatible with the adopted models \citep{luh99,luh03},
dwarf colors and bolometric corrections \citep{leg92,luh99}, and a distance
modulus of 5.76 \citep{wic98}. The plateau near $H=13.5$ in the resulting 
isochrone in Figure~\ref{fig:color}, which does not appear in a plot of the
isochrone in $T_{\rm eff}$ versus $L_{\rm bol}$, reflects the rapid increase 
in $I-K_s$ between M5 and M9.
Because standard values of intrinsic $I-Z$ as a function of spectral type
are not available, we cannot plot the isochrone in $I-Z$ versus $H$. 
However, the lower envelope of the sequence of known members in $I-Z$ versus 
$H$ effectively delineates the location of the 10~Myr isochrone 
(except at late types, or $I-Z>1.7$) 
and the line above which we should search for new members.
Approximately 23 stars lack spectra and fall near the member sequences 
in both of the CMDs in Figure~\ref{fig:color}. 
In the next section, we describe spectroscopy of 22 of these stars, which are 
listed in Tables~\ref{tab:back} and \ref{tab:mem}. The remaining source 
is 2MASSs~0435283+241000, which has dereddened photometry of $H=15.75$, 
$I-Z=2.33$, and $I-K_s=4.89$.
We did not observe this star because it is at the detection limits of both the
optical and 2MASSs data and therefore its positions in the CMDs 
are highly uncertain.

\section{Spectroscopic Observations of Candidates}
\label{sec:obs}

Table~\ref{tab:log} summarizes our observations of the 22 candidate members
of Taurus identified in the previous section. One of these objects, 
2MASSs~J0419012+280248, was observed on two occasions. 
The FAST spectrometer \citep{fab98}
on the 1.5~m Tillinghast reflector at the Fred Lawrence Whipple Observatory
was operated with the 300~l~mm$^{-1}$
grating ($\lambda_{\rm blaze}=4800$~\AA) and $2\farcs0$
slit, providing a resolution of FWHM$=5$~\AA.
The Keck~I low-resolution imaging spectrometer (LRIS; Oke et al.\ 1995) was 
configured with the 400~l~mm$^{-1}$ grating ($\lambda_{\rm blaze}=8500$~\AA), 
GG495 blocking filter, and $1\farcs0$ slit (FWHM$=6$~\AA).
With the Blue Channel spectrometer at the MMT Observatory,
we used the 600~l~mm$^{-1}$ grating ($\lambda_{\rm blaze}=9630$~\AA), LP495 
blocking filter, and $1\farcs0$ slit (FWHM$=2.7$~\AA).
Each spectrum was collected with the slit rotated to the parallactic angle. 
The exposure times ranged from 300 to 2700~s. 
After bias subtraction and flat-fielding,
the spectra were extracted and calibrated in wavelength with arc lamp data.
The spectra were then corrected for the sensitivity functions of the detectors, 
which were measured from observations of spectrophotometric standard stars. 

\section{Spectral Classification of Candidates}
\label{sec:class}

We measure spectral types and assess membership for the 22 candidate Taurus
members that were observed spectroscopically by applying the methods of
classification described in our previous studies of Taurus and other young 
populations \citep{luh99,bri02}. 

\begin{figure}
\epsscale{0.45}
\plotone{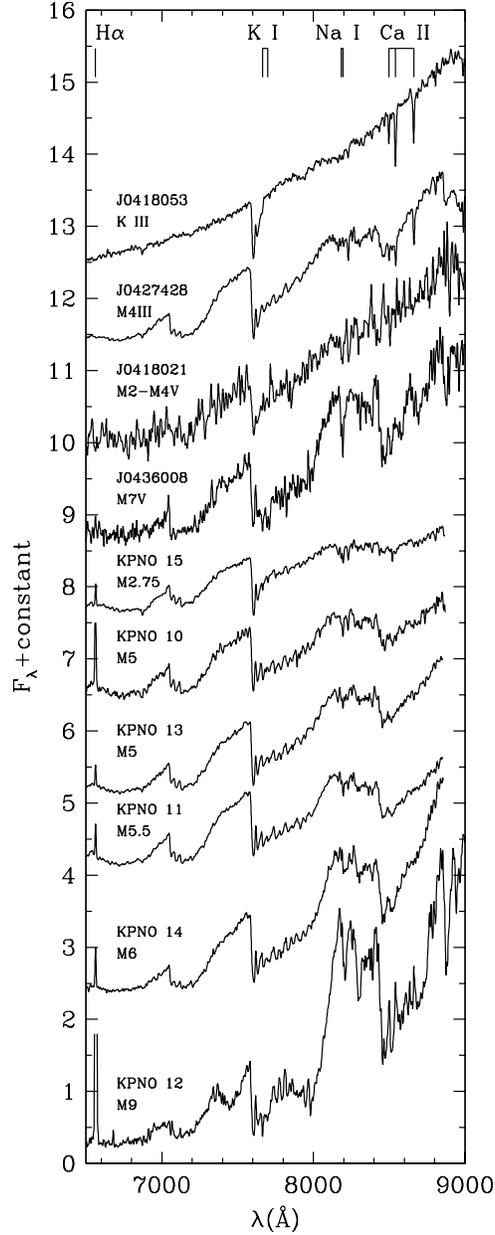}
\caption{
Spectra of candidate members of the Taurus star-forming region.
The spectra of 2MASSs~J0418053+282801
and J0427428+262256 exhibit the reddening and strong Ca~II absorption that 
are expected of background field giants.  
The strong K~I and Na~I absorption in the data for
2MASSs~J0436008+225517 is indicative of a field dwarf.
Because this star and 2MASSs~J0418021+281748 fall below the main sequence for 
the distance of Taurus and show significant reddening, they are probably 
background field dwarfs.
The remaining six objects are confirmed as pre-main-sequence sources by 
the weak absorption in K~I and Na~I and strong emission in H$\alpha$. 
All data are smoothed to a resolution of 8~\AA\ and normalized at 7500~\AA.
}
\label{fig:spec1}
\end{figure}

We classify 13 candidates as background giants and one candidate as a 
background early-type star. Strong absorption in the Ca~II triplet is the most 
distinguishing feature of the giants, as illustrated for two of these stars in 
Figure~\ref{fig:spec1}. 
The data for the other eight candidates are displayed in Figure~\ref{fig:spec1}.
Two of these stars, 2MASSs~J0418021+281748 and J0436008+225517, are below the 
main sequence when placed on the H-R diagram with the distance of Taurus, which 
indicates that they are probably field dwarfs behind the star-forming region.
Although young stars that are observed in scattered light can appear at 
comparable positions on the H-R diagram (\citet{bri02}, \S~\ref{sec:ext}), such 
stars usually exhibit emission in H$\alpha$ or some other evidence of youth. 
However, no such signatures are found in the data for these two stars. 
In addition, the strength of the K~I and Na~I absorption in the spectrum of 
2MASSs~J0436008+225517 provides conclusive evidence that it is a field dwarf.

\begin{figure}
\epsscale{0.9}
\plotone{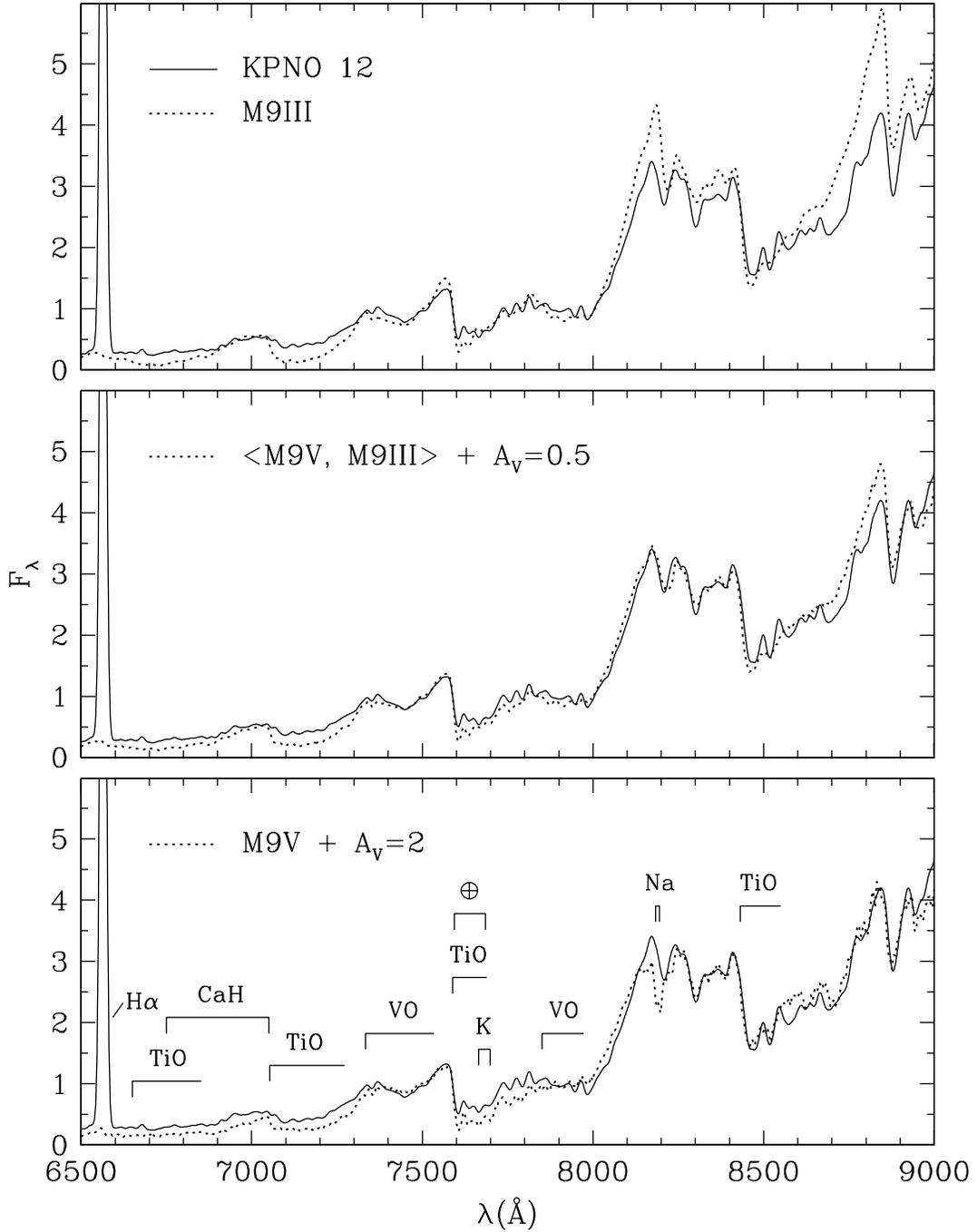}
\caption{
Spectrum of the new Taurus member KPNO-Tau~12 compared to 
data for M9~III, M9~V, and an average of the two types. Reddening is applied
to the latter data to optimize the matches to the spectrum of KPNO-Tau~12.
All data are smoothed to a resolution of 18~\AA\ and normalized at 7500~\AA.
}
\label{fig:spec2}
\end{figure}

The spectra for the six remaining sources in Figure~\ref{fig:spec1}
exhibit evidence of youth, and thus membership in Taurus, in the form of 
the weak K~I and Na~I absorption features that are characteristic of 
pre-main-sequence objects \citep{mar96,luh98a,luh98b,luh99}. 
The membership of five of these sources is independently established by 
the presence of reddening in their spectra and their positions above the main 
sequence for the distance of Taurus, which indicate that they cannot be field
dwarfs in the foreground or the background of the cloud, respectively.
The equivalent widths of H$\alpha$ emission in KPNO-Tau~11, 13, and 15
are consistent with those of both active field dwarfs and young objects. 
Meanwhile, the much stronger emission in KPNO-Tau~10, 12, and 14 is common
among the latter but rare in the former \citep{giz00}.
For the coolest new member, KPNO-Tau~12, we find that the best match 
to the optical spectrum is provided with an average of dwarf and giant spectra 
(normalized at 7500~\AA), as shown in Figure~\ref{fig:spec2}, which is 
consistent with the results of our previous classifications of young objects 
later than M6 \citep{luh99,bri02}. The spectrum for KPNO-Tau~12 in
Figure~\ref{fig:spec2} is from the November observations with Keck. 
We did not use the January data from the MMT in measuring
the spectral type of this object because the signal-to-noise was too low.

The classifications and other measurements for the 16 background field stars
and the six new Taurus members are listed in Tables~\ref{tab:back} and
\ref{tab:mem}, respectively. The photometric errors are $\sim0.1$~mag for 
KPNO-Tau~12 and $\sim0.03$~mag for the remaining sources.

\section{Properties of New Members}
\label{sec:prop}

In this section, we use the photometry, spectroscopy, and theoretical
evolutionary models to examine the physical properties of the six new Taurus
members. 

\subsection{Disk and Accretion Signatures}

\begin{figure}
\epsscale{0.9}
\plotone{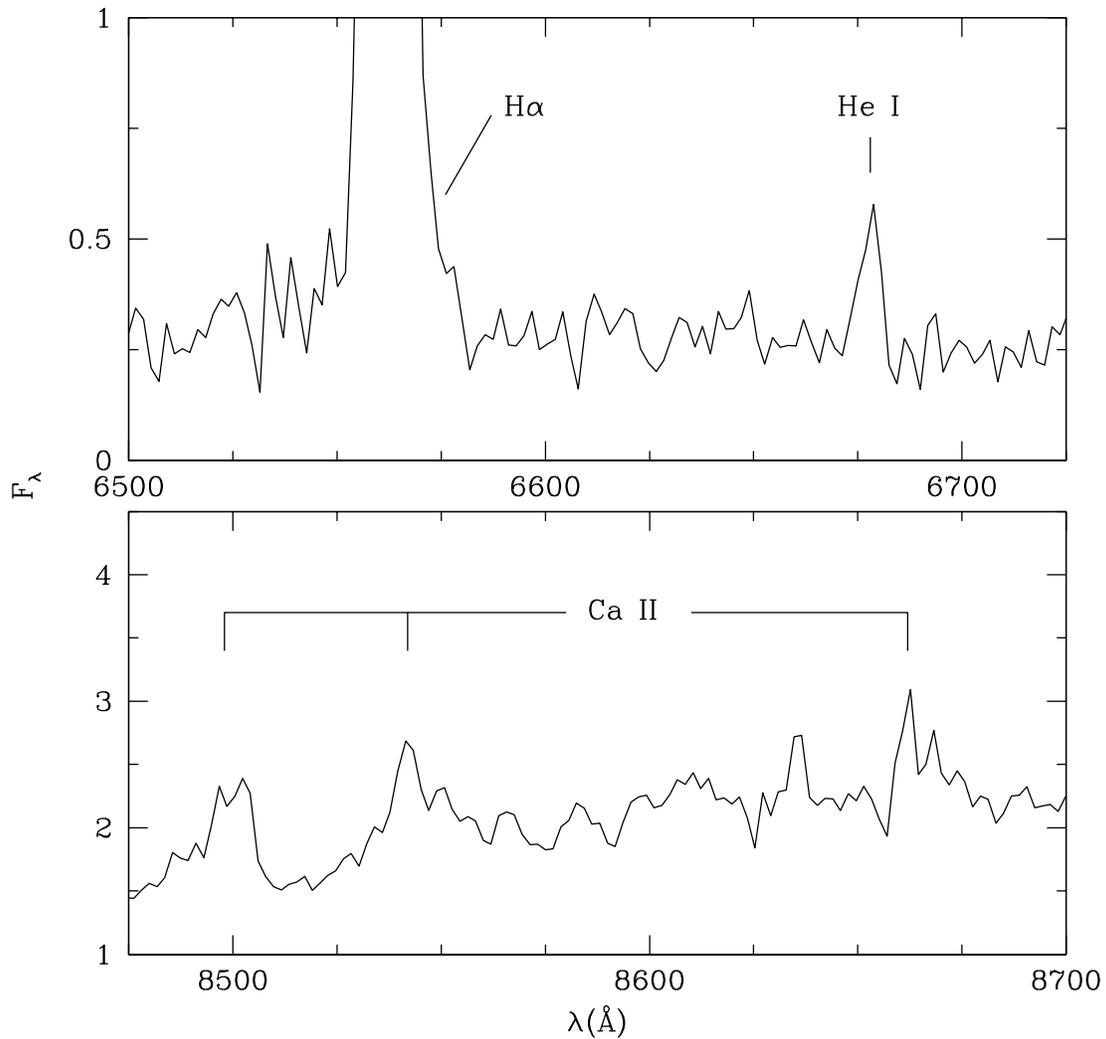}
\caption{
Emission lines in the spectrum of the new Taurus member KPNO-Tau~12.
At a mass of $\sim0.02$~$M_\odot$, this is the least massive object observed 
to date with He~I and Ca~II emission lines, which are typically found in stars
undergoing intense accretion. These data have a resolution of 6~\AA.
}
\label{fig:spec3}
\end{figure}

Excess continuum emission and optical emission lines are signatures of 
circumstellar disks and accretion that could appear in the available data for
the new Taurus members.
The near-IR colors of the six new members match those of 
reddened dwarfs and have no significant excess emission in the $K$-band.
This is consistent with the observation that only half of the known members 
of Taurus exhibit $K$-band excess emission, while a much larger fraction show 
evidence for disks in photometry at longer bandpasses such as $L$
\citep{kh95,hai00}. 
Five of the new members show moderate to strong emission in 
H$\alpha$ ($W_{\lambda}=6$-36~\AA) that is typical of young stars in Taurus,
while the M9 object KPNO-Tau~12 exhibits a far larger equivalent 
width. A given value of $L_{H\alpha}/L_{\rm bol}$ corresponds to larger
equivalent widths at later spectral types as the stellar continuum 
surrounding H$\alpha$ becomes weaker. In addition, when this faint continuum
is detected at low signal-to-noise, it is possible to erroneously measure
arbitrarily large values of the equivalent width. However, in the case of
KPNO-Tau~12, the continuum near H$\alpha$ is well-detected,
as demonstrated in Figure~\ref{fig:spec3}. In measuring the equivalent
width of H$\alpha$, the continuum level across the line was
taken to be the level from the best fit M9 standard (\S~\ref{sec:class}) 
after normalizing the two spectra at the surrounding continuum 
(6400-6500 and 6600-6700~\AA). In this way, the equivalent width of H$\alpha$ 
is confidently measured to be between 850 and 1050~\AA\ in the November 
spectrum. Because of insufficient signal-to-noise in the continuum surrounding
H$\alpha$, an equivalent width was not measured from the January spectrum.

To compare the H$\alpha$ emission in KPNO-Tau~12 to that 
in other M-type sources, we consider the ratio $L_{H\alpha}/L_{\rm bol}$.
We arrived at two separate flux calibrations for the November spectrum of 
KPNO-Tau~12 by using the $I$-band photometry of this source 
from \citet{bri02} and the data for the spectrophotometric standard. We adopted 
the average of these calibrations, which differed by 20\%. From the resulting
spectrum, we measured an H$\alpha$ flux of 
$2.45\pm0.25\times10^{-15}$~erg~s$^{-1}$~cm$^{-2}$ in the November
data. Meanwhile, the flux of H$\alpha$ in the January spectrum was 30\% higher.
After correcting the November measurement for an extinction of $A_V=0.5$ 
and combining the result with the bolometric luminosity derived in the next 
section, we find log~$(L_{H\alpha}/L_{\rm bol})=-2.6\pm0.12$, which 
is slightly higher than the strongest emission observed in late-M dwarfs
\citep{sch91,mar99,lie99} and two orders of magnitude greater than the typical 
values for these objects \citep{giz00}.
The only known young late-M sources with comparable H$\alpha$ emission are 
LS-RCrA~1 in R Coronae Australis \citep{fer01} and S~Ori~71 in $\sigma$~Orionis 
\citep{bar02}. 
Several other young low-mass objects have been found with 
H$\alpha$ intensities that are only a few times lower than those in 
KPNO-Tau~12, LS-RCrA~1, and S~Ori~71 \citep{zap02,bri02,luh03}. 

There are additional indications that active accretion is probably occurring
in KPNO-Tau~12. The TiO bands at 6600-7300~\AA\ are weaker than 
those of the best fit standard spectrum in Figure~\ref{fig:spec2}, which is 
suggestive of veiling from blue excess continuum emission. On the other hand, 
this behavior 
could imply a spectral type later than our classification of M9 since the TiO 
bands become weaker and eventually disappear from M to L types. But given the 
good quality of the fit of M9 to the data beyond 7300~\AA\ and the presence of 
intense H$\alpha$ emission, the weak appearance of the TiO bands is probably
a result of continuum veiling. We also detect emission in He~I at 
6678~\AA\ and in the Ca~II IR triplet, as shown in Figure~\ref{fig:spec3}.
We measure equivalent widths of $8\pm2$~\AA\ for He~I and $4\pm1$, $2.5\pm1$, 
and $1.5\pm0.5$~\AA\ for Ca~II at 8498, 8542, and 8662~\AA.
KPNO-Tau~12 is the coolest young object known to exhibit these emission lines,
which are usually indicators of intense accretion when observed in young 
stars \citep{muz98,ber01}.
Alternatively, emission in He~I and Ca~II, as well as strong H$\alpha$ and 
continuum veiling, can be attributed to a magnetic flare event, as in the case 
of the field M9.5 dwarf 2MASSW~J0149090+295613 \citep{lie99}. 
However, since the duty cycle of such flares is low in field objects
\citep{giz00,lie03} and the emission in KPNO-Tau~12 was strong at both epochs 
of our observations, the emission in KPNO-Tau~12 is probably not from a flare 
event. Given that KPNO-Tau~12 is member of a star-forming region, it is much 
more likely that the emission in H$\alpha$, He~I, Ca~II, and the blue continuum 
are the result of active accretion.

\subsection{Extinctions, Temperatures, and Luminosities}
\label{sec:ext}

\begin{figure}
\epsscale{0.85}
\plotone{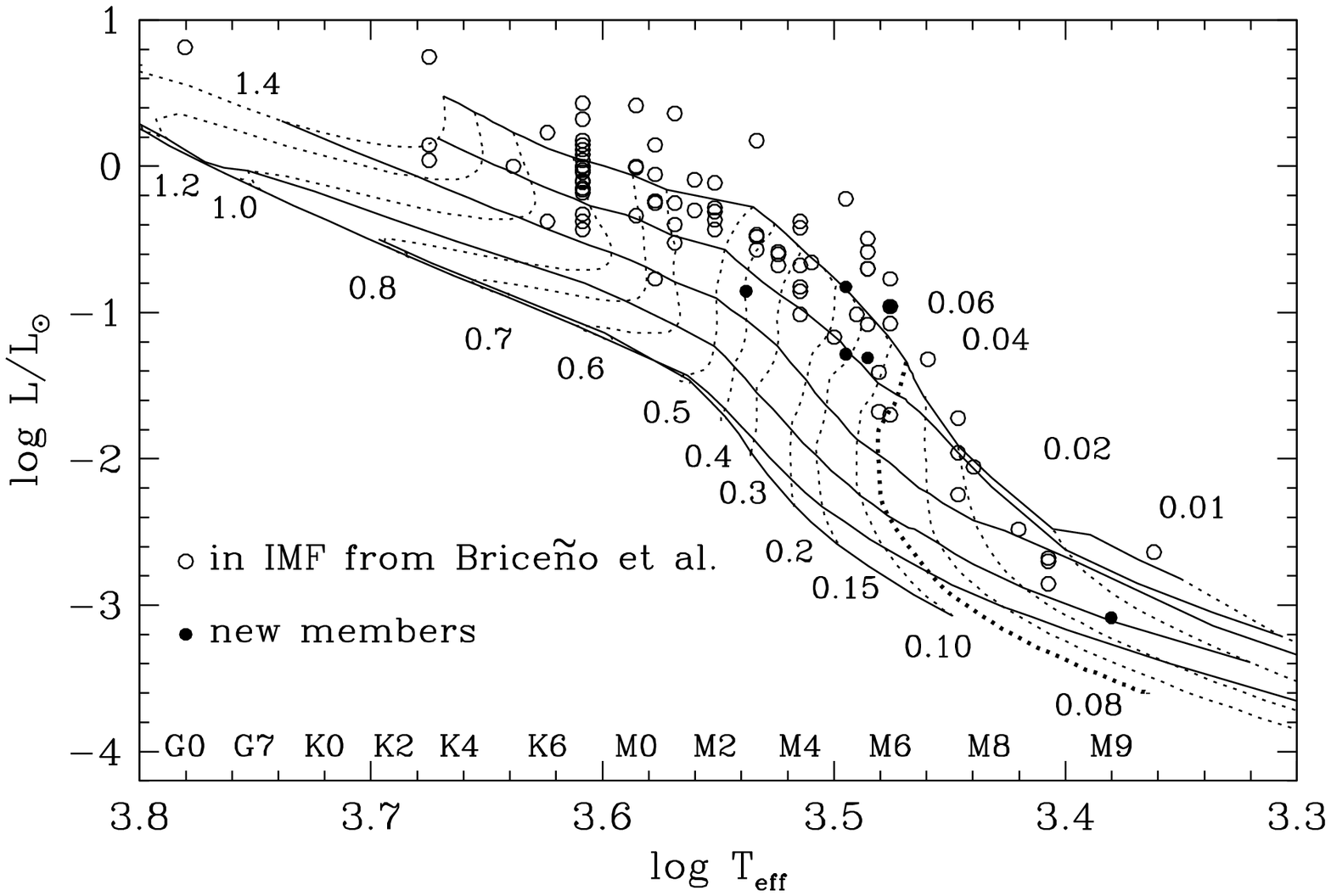}
\caption{
H-R diagram for young objects in the IMF for Taurus reported by \citet{bri02}
({\it open circles}). This sample is 
extinction-limited ($A_V\leq4$) and applies to 8.4~deg$^2$ of Taurus
surveyed by \citet{bri98}, \citet{luh00}, and \citet{bri02}. 
We have spectroscopically identified six new members in these
fields ({\it solid points}), which have $A_V\leq4$ and thus are added to the IMF
from \citet{bri02} in Figure~\ref{fig:imf}.
The theoretical evolutionary models of \citet{bar98} are shown, where
the horizontal solid lines are isochrones representing ages of 1, 3, 10, 30,
and 100~Myr and the main sequence, from top to bottom.
The M spectral types have been converted to effective temperatures with a
scale such that GG~Tau~Ba and Bb fall on the same model isochrone as
Aa and Ab and that the M8-M9 members of Taurus and IC~348 have
model ages that are similar to those of the earlier members \citep{luh03}.
}
\label{fig:hr}
\end{figure}

Following the procedures described by \citet{bri02}, we estimate
extinctions, effective temperatures, and bolometric luminosities from
the spectral types and photometry for the six new Taurus members.
The values for these parameters are listed in Table~\ref{tab:mem}.
The combined uncertainties in $A_J$, $J$, and BC$_J$ ($\sigma\sim0.14$, 0.03, 
0.1) correspond to errors of $\pm0.07$ in the relative values of 
log~$L_{\rm bol}$.
When an uncertainty in the distance modulus is included ($\sigma\sim0.2$), 
the total uncertainties are $\pm0.11$. We use the temperatures and 
luminosities in Table~\ref{tab:mem} to place the six sources on the 
H-R diagram in Figure~\ref{fig:hr}. For reference, we also show the 
previously known Taurus members that have $A_V\leq4$ and are within the
8.4~deg$^2$ surveyed by \citet{bri98}, \citet{luh00}, and \citet{bri02}, 
which comprise the IMF sample from \citet{bri02}.
The five new members at M3-M6 fall within this sequence of members,
which is spread between $<1$ and 10~Myr for most masses.
Meanwhile, the new M9 source KPNO-Tau~12 appears near the 30~Myr
model isochrone of \citet{bar98}, as does one
of the previously known M8.5 members. An age of 30~Myr is much larger than
expected given the median age of 1~Myr for the Taurus population and the
apparent presence of active accretion in KPNO-Tau~12.
Interestingly, the one other young late-M source known to exhibit accretion 
signatures in the form of He~I and Ca~II emission, 
LS-RCrA~1, also falls much lower on the H-R diagram than expected \citep{fer01}.

We consider three possible explanations for the anomalously old ages implied
by the H-R diagrams of some late-M sources, particularly KPNO-Tau~12
and LS-RCrA~1. First, there could be deficiencies in either
the models or the conversion of spectral types to temperatures.
In this work, we use a temperature scale that has been adjusted at late spectral
types so that the low-mass members of Taurus and IC~348 have median ages 
comparable to the ones of the higher mass sources \citep{luh03}. As a result, 
the adopted combination of models and temperature scale should produce the 
most reliable age and mass estimates that are currently possible. 
However, as suggested by \citet{fer01} in the case of LS-RCrA~1, intense 
accretion could produce an evolutionary path on the H-R diagram that differs
from that predicted by models of non-accreting objects. Indeed, both of
the known late-M objects with evidence for accretion in the form
of He~I and Ca~II emission are subluminous on the H-R diagram.
\citet{fer01} proposed a second explanation for the low luminosity of 
LS-RCrA~1 in which this object is seen in scattered light. For a young star 
that is occulted by an optically thick structure such as an edge-on disk, the
observed photometry measures only the scattered light. As a result, the 
luminosity implied by the photometry is an underestimate, and the star 
appears subluminous for its temperature on the H-R diagram. 
For a low-mass source undergoing intense accretion, it is possible
that a significant fraction of the object is blocked by accreting circumstellar
material, which could account for the correlation between accretion activity
and low luminosity for KPNO-Tau~12 and LS-RCrA~1.
However, this occulting mechanism would need to be more efficient for brown 
dwarfs than for stars since young stars in Taurus with He~I emission do not 
have systematically lower luminosities than other stellar members.
As a third possibility, the old ages inferred from the model isochrones
could result from errors in the luminosity estimates of young sources.
The finite width of the Taurus sequence on the H-R diagram could arise 
not from a spread in ages but from 
various sources such as extinction uncertainties, unresolved binaries, 
variability from accretion and from rotation of spotted surfaces, and 
differences in distances to individual members \citep{kh90,har01}. 
If so, then the width of the sequence in log~L should remain roughly constant
at all masses, and thus correspond to an increasing apparent spread of ages as 
the separation between model isochrones in log~L decreases from stellar to 
substellar masses. This phenomenon probably produces erroneously old ages on 
the model isochrones for at least some low-mass members of young clusters, but 
would not explain the fact that both late-type objects showing He~I emission
are subluminous.

\subsection{Masses}

We now consider the masses of the new Taurus members. By combining the
positions on the H-R diagram with the evolutionary models of \citet{bar98},
we infer masses of 0.10, 0.13, 0.15, 0.16, and 0.4~$M_{\odot}$ for 
KPNO-Tau~14, 11, 13, 10, and 15.
These mass estimates are not sensitive to uncertainties in the luminosities
since the predicted evolution of a star is mostly vertical on the H-R diagram 
for ages of $<10$~Myr. However, this is not true at substellar masses, 
where the mass tracks have significant components in both temperature and 
luminosity. The models imply a mass of 0.03~$M_{\odot}$ for the 
M9 object KPNO-Tau~12 based on its position on the
H-R diagram. Under two of the three explanations for the low luminosity of 
this source (\S~\ref{sec:ext}), the luminosity estimate is uncertain. 
In these cases, a more reliable mass is derived by placing 
the object on the 1~Myr isochrone for the adopted temperature, which would
imply a mass of 0.015~$M_{\odot}$ from the models of \citet{bar98}. 
For the third explanation in which the temperature and luminosity are 
significantly affected by accretion, KPNO-Tau~12 could
have an even lower mass \citep{fer01}.  
If this object does have a mass below 0.015~$M_{\odot}$, one might 
be tempted to refer to it as an accreting free-floating planetary mass object.
However, the various properties of low-mass brown dwarfs -- such as 
the evidence of intense accretion in objects like KPNO-Tau~12 -- indicate that 
they probably form in a star-like fashion rather than in circumstellar disks 
\citep{bri02}. Therefore, the term ``planetary" has little or no applicability. 
For the purposes of this work, we adopt a mass of 0.02~$M_{\odot}$ for this 
object.

\begin{figure}
\epsscale{1}
\plotone{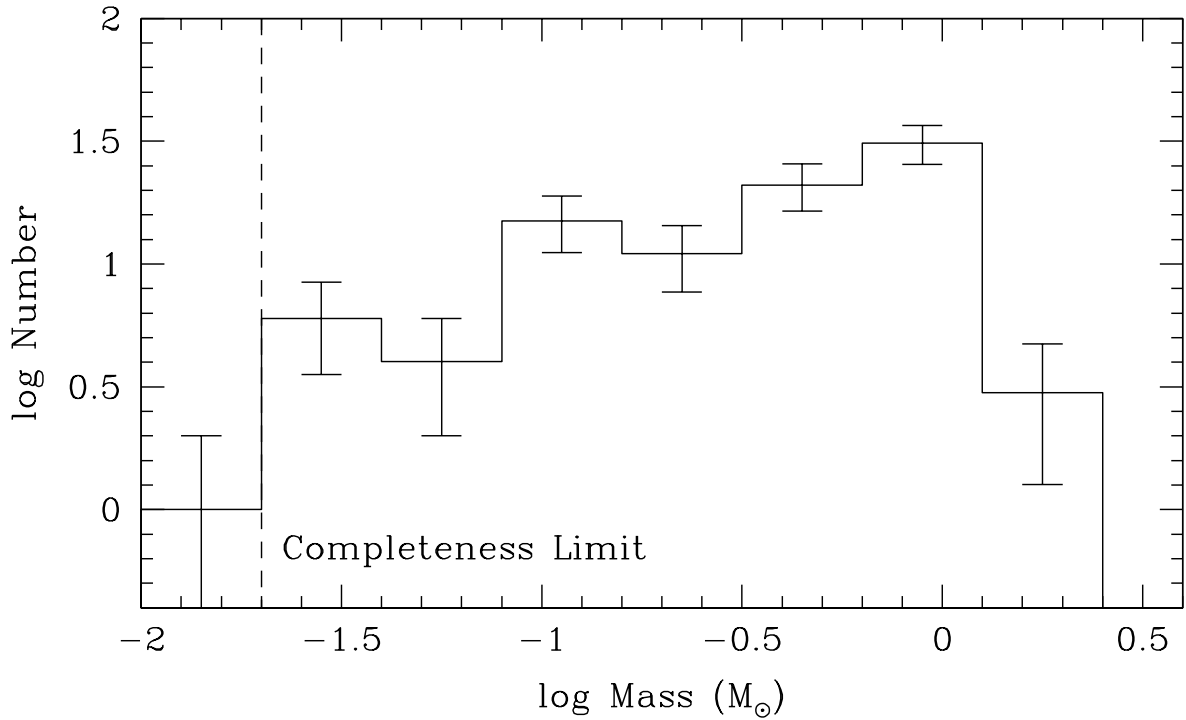}
\caption{
IMF for an extinction-limited sample ($A_V\leq4$) of young objects in the 
8.4~deg$^2$ of Taurus surveyed by \citet{bri98}, \citet{luh00}, and 
\citet{bri02}. This sample consists of the 86 Taurus members in the IMF
from \citet{bri02} and the six new members found in this work.
In the units of this diagram, the Salpeter slope is 1.35.
}
\label{fig:imf}
\end{figure}

As mentioned previously, the IMF presented by \citet{bri02} included the 86
known Taurus members that fell within the 8.4~deg$^2$ surveyed by 
\citet{bri98}, \citet{luh00}, and \citet{bri02} and that had extinctions of 
$A_V\leq4$. There remained a few potential members that lacked spectroscopy
within these defining parameters, so the IMF sample was not 100\% complete.
However, because the candidates were not concentrated at a particular 
magnitude range, \citet{bri02} concluded that the sample of confirmed 
members was unbiased in mass down to the completeness limit of the photometry,
which corresponded to 0.02~$M_{\odot}$ for $A_V=4$.
Through spectroscopy of those candidates, we have found six new members, 
all of which have $A\leq4$. 
After adding these sources to the extinction-limited sample from \citet{bri02},
we arrive at the IMF in Figure~\ref{fig:imf}, which should be virtually
100\% complete for $A_V\leq4$ and masses above 0.02~$M_{\odot}$ in the
8.4~deg$^2$ survey fields. Because of the small number of new members that are
added to the IMF from \citet{bri02} and the wide range of masses of these new 
sources, the shapes of the IMFs here and in \citet{bri02} do not differ 
significantly. Thus, the conclusions from \citet{bri02} regarding the observed 
variations in the IMFs of Taurus and the Trapezium remain unchanged.

\section{Conclusions}

We have obtained spectroscopy for candidate young stars and brown dwarfs that 
appeared in an optical and IR photometric survey of 8~deg$^2$ of the Taurus 
star-forming region by \citet{bri02}.
We have estimated individual extinctions for the stars detected in 
the survey and have plotted the dereddened photometry on color-magnitude 
diagrams. Within an extinction limit of $A_V\leq8$, we identified 22 sources
that were potential members of Taurus and that lacked previous spectroscopy.
Through optical spectroscopy of these candidates, we have discovered six
new members with spectral types ranging from M2.75 to M9.
From the spectral types and photometry of these objects, we have estimated
extinctions, effective temperatures, and bolometric luminosities, and have
combined these results with the evolutionary models of \citet{bar98}
to infer individual masses that range from 0.4 to 0.02~$M_\odot$.
These new members fall within the extinction limit 
of $A_V\leq4$ that defined the sample from which \citet{bri02} derived an IMF.
Therefore, we have added them to that extinction-limited sample and have 
presented the updated IMF, which applies to the 8.4~deg$^2$ in Taurus surveyed 
by \citet{bri98}, \citet{luh00}, and \citet{bri02}. 
The new M9 source is a particularly interesting discovery from this work.
It has the second latest spectral type of the known members of Taurus.
In addition, it exhibits the strongest H$\alpha$ emission observed to date for 
a late-M source ($W_{\lambda}\sim950$~\AA, log~$(L_{H\alpha}/L_{\rm bol})=-2.6$)
and is the least massive object found with emission in He~I 6678~\AA\ and 
the Ca~II IR triplet, which are suggestive of active accretion. High-resolution
spectroscopy of the emission lines in this object would provide valuable 
insight into the formation of low-mass brown dwarfs.

\acknowledgements

We thank Perry Berlind and Mike Calkins for performing the FAST observations.
K. L. was supported by grant NAG5-11627 from the NASA Longterm Space 
Astrophysics program. C. B. received partial support from grant S1-2001001144 
of the Fondo Nacional de Ciencia y Tecnolog{\'\i}a (FONACYT) of Venezuela.
We are grateful to France Allard and Isabelle Baraffe for 
access to their most recent calculations.  
This research has made use of the NASA/IPAC Infrared Science Archive, which is
operated by the Jet Propulsion Laboratory, California Institute of Technology, 
under contract with the National Aeronautics and Space Administration. 
Some of the data presented herein were obtained at the W. M. Keck Observatory, 
which is operated as a scientific partnership among the California Institute 
of Technology, the University of California, and the National Aeronautics and 
Space Administration.  The Observatory was made possible by the generous 
financial support of the W. M. Keck Foundation.
We wish to extend special thanks to those of Hawaiian ancestry on whose sacred
mountain we are privileged to be guests. Without their generous hospitality,
some of the observations presented herein would not have been possible.
Some of the data in this work were obtained at the MMT Observatory,
a joint facility of the Smithsonian Institution and the University of Arizona.

\newpage

\begin{deluxetable}{llllrrrrr}
\tablewidth{0pt}
\tablecaption{Background Stars \label{tab:back}}
\tablehead{
\colhead{2MASSs} &
\colhead{$\alpha$(J2000)\tablenotemark{a}} &
\colhead{$\delta$(J2000)\tablenotemark{a}} &
\colhead{Spectral Type} &
\colhead{$I$\tablenotemark{b}} &
\colhead{$I-Z$\tablenotemark{b}} &
\colhead{$J-H$\tablenotemark{a}} & \colhead{$H-K_s$\tablenotemark{a}}
& \colhead{$K_s$\tablenotemark{a}} \\
\colhead{} &
\colhead{h m s} &
\colhead{$\arcdeg$ $\arcmin$ $\arcsec$} &
\colhead{} &
\colhead{} &
\colhead{} &
\colhead{} &
\colhead{} &
\colhead{}
}
\startdata
J0413179+281143 & 04 13 17.97 & 28 11 43.0 & giant & 14.55 & 1.42 & 1.19 & 0.49 & 10.22 \\
J0417519+282551 & 04 17 51.94 & 28 25 51.1 & giant & 15.37 & 1.39 & 1.04 & 0.56 & 11.18 \\
J0418021+281748 & 04 18 02.15 & 28 17 48.9 & M2-M4V & 19.49 & 1.93 & 1.12 & 0.80 & 14.13 \\
J0418053+282801 & 04 18 05.38 & 28 28 01.3 & giant & 16.08 & 1.75 & 1.59 & 0.72 & 10.33 \\
J0418423+281140 & 04 18 42.38 & 28 11 40.7 & giant & 12.01 & 1.22 & 0.65 & 0.29 & 9.28 \\
J0419273+281301 & 04 19 27.37 & 28 13 01.1 & giant & 13.11 & 1.51 & 1.55 & 0.55 & 8.16 \\
J0426255+260653 & 04 26 25.50 & 26 06 53.4 & giant & 15.46 & 1.28 & 1.00 & 0.48 & 11.08 \\
J0426344+260740 & 04 26 34.45 & 26 07 40.2 & giant & 15.85 & 1.72 & 1.32 & 0.70 & 10.75 \\
J0427428+262256 & 04 27 42.87 & 26 22 56.7 & giant & 13.02 & 1.57 & 1.42 & 0.61 & 7.33 \\
J0432113+261323 & 04 32 11.33 & 26 13 23.6 & giant & 13.64 & 1.73 & 1.09 & 0.40 & 9.24 \\
J0432138+263046 & 04 32 13.82 & 26 30 46.1 & giant & 13.33 & 1.38 & 1.30 & 0.45 & 9.25 \\
J0433080+255643 & 04 33 08.03 & 25 56 43.7 & giant & 13.97 & 1.67 & 1.49 & 0.61 & 8.67 \\
J0433293+261809 & 04 33 29.33 & 26 18 09.6 & giant & 15.10 & 1.45 & 1.38 & 0.58 & 10.29 \\
J0433341+181426 & 04 33 34.11 & 18 14 26.2 & giant & 15.13 & 1.29 & 0.89 & 0.34 & 11.39 \\
J0433513+262614 & 04 33 51.33 & 26 26 14.3 & early & 16.58 & 1.67 & 1.14 & 0.65 & 11.84 \\
J0436008+225517 & 04 36 00.85 & 22 55 17.4 & M7V$\pm0.5$ & 19.92 & 2.04 & 0.60 & 0.37 & 15.69 \\
\enddata
\tablenotetext{a}{2MASS Spring 1999 Release Point Source Catalog.}
\tablenotetext{b}{\citet{bri02}.}
\end{deluxetable}

\begin{deluxetable}{lllllrrrrrrrr}
\tabletypesize{\scriptsize}
\rotate
\tablewidth{0pt}
\tablecaption{New Members of Taurus \label{tab:mem}}
\tablehead{
\colhead{ID} &
\colhead{2MASSs ID} &
\colhead{$\alpha$(J2000)\tablenotemark{a}} & 
\colhead{$\delta$(J2000)\tablenotemark{a}} &
\colhead{Spectral Type/$W_{\lambda}$(H$\alpha$)} &
\colhead{$T_{\rm eff}$\tablenotemark{b}} &
\colhead{$A_J$} & \colhead{$L_{\rm bol}$} & 
\colhead{$I$\tablenotemark{c}} & \colhead{$I-Z$\tablenotemark{c}} &
\colhead{$J-H$\tablenotemark{a}} & \colhead{$H-K_s$\tablenotemark{a}} & 
\colhead{$K_s$\tablenotemark{a}} \\
\colhead{} &
\colhead{} &
\colhead{h m s} &
\colhead{$\arcdeg$ $\arcmin$ $\arcsec$} &
\colhead{} &
\colhead{} &
\colhead{} &
\colhead{} &
\colhead{} &
\colhead{} &
\colhead{} &
\colhead{} &
\colhead{}
}
\startdata
KPNO-Tau 10 & J0417495+281331  &    04 17 49.54 &    28 13 32.0 &       M5$\pm$0.25/36$\pm$5 &   3125 &    0.14 &   0.052 &   13.94 &     1.25 &     0.77 &     0.32 &    10.78  \\
KPNO-Tau 11 & J0418302+274320  &    04 18 30.30 &    27 43 20.6 &     M5.5$\pm$0.25/14$\pm$1 &   3058 &    0.00 &   0.049 &   13.72 &     1.22 &     0.61 &     0.23 &    11.01  \\
KPNO-Tau 12 & J0419012+280248  &    04 19 01.26 &    28 02 48.7 &   M9$\pm$0.25/950$\pm$100 &   2400 &    0.14 & 0.00082 &   19.69 &     2.04 &     0.82 &     0.50 &    14.94  \\
KPNO-Tau 13 & J0426573+260628  &    04 26 57.31 &    26 06 28.8 &       M5$\pm$0.25/10$\pm$1 &   3125 &    0.70 &    0.15 &   13.70 &     1.57 &     1.11 &     0.56 &     9.60  \\
KPNO-Tau 14 & J0433078+261606  &    04 33 07.81 &    26 16 06.6 &      M6$\pm$0.25/40$\pm$10 &   2990 &    0.85 &    0.11 &   14.98 &     1.86 &     1.10 &     0.54 &    10.27  \\
KPNO-Tau 15 & J0435510+225240  &    04 35 51.10 &    22 52 40.1 & M2.75$\pm$0.25/5.8$\pm$0.5 &   3451 &    0.56 &    0.14 &   13.63 &     1.48 &     0.98 &     0.33 &    10.01  \\
\enddata
\tablenotetext{a}{2MASS Spring 1999 Release Point Source Catalog.}
\tablenotetext{b}{Temperature scale of \citet{luh03}.}
\tablenotetext{c}{\citet{bri02}.}
\end{deluxetable}

\begin{deluxetable}{lll}
\tablewidth{0pt}
\tablecaption{Observing Log \label{tab:log}}
\tablehead{
\colhead{Date} &
\colhead{Telescope + Instrument} &
\colhead{2MASSs}
}
\startdata
2002 Jan 12 & MMT + Blue Channel & J0419012+280248 \\
2002 Sep 12 & FLWO 1.5 m + FAST & J0418423+281140 \\
\nodata & \nodata & J0432113+261323 \\
\nodata & \nodata & J0435510+225240 \\
\nodata & \nodata & J0418302+274320 \\
2002 Sep 27 & FLWO 1.5 m + FAST & J0417495+281331 \\
2002 Nov 5 & Keck + LRIS & J0436008+225517 \\
\nodata & \nodata & J0419012+280248 \\
2002 Nov 10 & MMT + Blue Channel & J0418021+281748 \\
2002 Dec 7 & FLWO 1.5 m + FAST & J0413179+281143 \\
\nodata & \nodata & J0426573+260628 \\
\nodata & \nodata & J0432138+263046 \\ 
\nodata & \nodata & J0433078+261606 \\
\nodata & \nodata & J0433080+255643 \\
2002 Dec 14 & MMT + Blue Channel & J0419273+281301 \\
\nodata & \nodata & J0426344+260740 \\
\nodata & \nodata & J0427428+262256 \\
\nodata & \nodata & J0433293+261809 \\
\nodata & \nodata & J0433513+262614 \\
2002 Dec 15 & MMT + Blue Channel & J0417519+282551 \\
\nodata & \nodata & J0418053+282801 \\
\nodata & \nodata & J0426255+260653 \\
\nodata & \nodata & J0433341+181426 \\
\enddata
\end{deluxetable}

\newpage

\clearpage

\clearpage

\clearpage

\clearpage

\clearpage

\newpage

\begin{thebibliography}{}

\bibitem[Baraffe et al.(1998)]{bar98}
Baraffe, I., Chabrier, G., Allard, F., \& Hauschildt, P. H. 1998, \aap, 337, 403

\bibitem[Barrado y Navascu\'es et al.(2002)]{bar02}
Barrado y Navascu\'es, D., Zapatero Osorio, M. R., Mart{\'\i}n, E. L., 
B\'{e}jar, V. J. S., Rebolo, R., \& Mundt, R. 2002, \aap, 393, L85

\bibitem[Beristain et al.(2001)]{ber01}
Beristain, G., Edwards, S., \& Kwan, J. 2001, \apj, 551, 1037

\bibitem[Brice\~no et al.(1998)]{bri98}
Brice\~{n}o, C., Hartmann, L., Stauffer, J., \& Mart{\'\i}n, E. L., 1998, \aj, 
115, 2074 

\bibitem[Brice\~no et al.(2002)]{bri02}
Brice\~{n}o, C., Luhman, K. L., Hartmann, L., Stauffer, J. R., \& Kirkpatrick, 
J. D. 2002, \apj, 580, 317

\bibitem[Fabricant et al.(1998)]{fab98}
Fabricant, D., Cheimets, P., Caldwell, N., \& Geary, J. 1998, \pasp, 110, 79

\bibitem[Fern\'{a}ndez \& Comer\'{o}n(2001)]{fer01}
Fern\'{a}ndez, M., \& Comer\'{o}n, F. 2001, \aap, 380, 264

\bibitem[Gizis et al.(2000)]{giz00}
Gizis, J. E., Monet, D. G., Reid, I. N., Kirkpatrick, J. D., Liebert, J., 
\& Williams, R. J. 2000, \aj, 120, 1085

\bibitem[Haisch et al.(2000)]{hai00}
Haisch, K. E., Lada, E. A., \& Lada, C. J. 2000, \aj, 120, 1396

\bibitem[Hartmann(2001)]{har01}
Hartmann, L. 2001, \aj, 121, 1030

\bibitem[Kenyon \& Hartmann(1990)]{kh90}
Kenyon, S. J., \& Hartmann, L. 1990, \apj, 349, 197

\bibitem[Kenyon \& Hartmann(1995)]{kh95}
Kenyon, S. J., \& Hartmann, L. 1995, \apjs, 101, 117

\bibitem[Kirkpatrick et al.(1997)]{kir97}
Kirkpatrick, J. D., Henry, T. J., \& Irwin, M. J. 1997, \aj, 113, 1421

\bibitem[Leggett(1992)]{leg92}
Leggett, S. K. 1992, \apjs, 82, 351

\bibitem[Liebert et al.(2003)]{lie03}
Liebert, J., Kirkpatrick, J. D., Cruz, K. L., Reid, I. N., Burgasser, A., 
Tinney, C. G., \& Gizis, J. E. 2003, \aj, 125, 343

\bibitem[Liebert et al.(1999)]{lie99}
Liebert, J., Kirkpatrick, J. D., Reid, I. N., \& Fisher, M. 1999, \apj, 519, 345

\bibitem[Luhman(1999)]{luh99}
Luhman, K. L. 1999, \apj, 525, 466

\bibitem[Luhman(2000)]{luh00} 
Luhman, K. L. 2000, \apj, 544, 1044

\bibitem[Luhman et al.(1998a)]{luh98a}
Luhman, K. L., Brice\~{n}o, C., Rieke, G. H., \& Hartmann, L. W. 1998a, \apj, 
493, 909

\bibitem[Luhman \& Rieke(1998)]{lr98}
Luhman, K. L., \& Rieke, G. H. 1998, \apj, 497, 354

\bibitem[Luhman et al.(1998b)]{luh98b}
Luhman, K. L., Rieke, G. H., Lada, C. J., \& Lada, E. A. 1998b, \apj, 508, 347

\bibitem[Luhman et al.(2003)]{luh03}
Luhman, K. L., et al. 2003, \apj, submitted

\bibitem[Mart{\'\i}n et al.(1999)]{mar99}
Mart{\'\i}n, E. L., Basri, G., \& Zapatero Osorio, M. R. 1999, \aj, 118, 1005

\bibitem[Mart{\'\i}n et al.(2001)]{mar01}
Mart{\'\i}n, E. L., Dougados, C., Magnier, E., M\'{e}nard, F., Magazz\`{u}, A.,
Cuilandre, J.-C., \& Delfosse, X. 2001, \apj, 561, L195

\bibitem[Mart{\'\i}n et al.(1996)]{mar96}
Mart{\'\i}n, E. L., Rebolo, R., \& Zapatero Osorio, M. R. 1996, \apj, 469, 706

\bibitem[Muzerolle et al.(1998)]{muz98}
Muzerolle, J., Hartmann, L., \& Calvet, N. 1998, \aj, 116, 455

\bibitem[Oke et al.(1995)]{oke95}
Oke, J.B., et al. 1995, \pasp, 107, 375

\bibitem[Schneider et al.(1991)]{sch91}
Schneider, D. P., Greenstein, J. L., Schmidt, M., \& Gunn, J. E. 1991, \aj,
102, 1180

\bibitem[Strom \& Strom(1994)]{ss94}
Strom, K. M., \& Strom, S. E. 1994, \apj, 424, 237

\bibitem[Wichmann et al.(1998)]{wic98}
Wichmann, R., Bastian, U., Krautter, J., Jankovics, I., \& Ruci\'nski, S. M. 
1998, \mnras, 301, L39

\bibitem[Zapatero Osorio et al.(2002)]{zap02}
Zapatero Osorio, M. R., B\'{e}jar, V. J. S., Mart{\'\i}n, E. L., 
Barrado y Navascu\'es, D., \& Rebolo, R. 2002, \apj, 569, L99

\end{thebibliography}
\end{document}